\documentclass[]{rQUF2e}

\usepackage{amsmath}
\usepackage{color}
\usepackage{float}

\begin{document}


\title{Partial correlation analysis:\\ Applications for financial markets}

\author{Dror Y. Kenett$^{\ast}$ ${\dag}$\thanks{$^\ast$These authors had equal contribution. Corresponding author
email: drorkenett@gmail.com}, Xuqing Huang$^{\ast}$ $\dag$\thanks{$^\ast$These authors had equal contribution. Corresponding author
email: eqing2700@gmail.com}, Irena Vodenska${\dag}$ ${\ddag}$,\\ Shlomo Havlin${\dag}$ ${\S}$ and H. Eugene Stanley${\dag}$\\
\affil{$\dag$Center for Polymer Studies and Department of Physics,
\\Boston University, Boston, MA 02215 USA\\
$\ddag$Administrative Sciences Department, Metropolitan College,\\ Boston University, Boston MA 02215 USA\\
$\S$Department of Physics,\\Bar-Ilan University, Ramat-Gan 52900 Israel} \received{\today} }

\maketitle

\begin{abstract}
The presence of significant cross-correlations between the synchronous time evolution of a pair of equity returns is a well-known empirical fact. The Pearson correlation is commonly used to indicate the level of similarity in the price changes  for a given pair of stocks, but it does not measure whether other stocks influence the relationship between them. To explore the influence of a third stock on the relationship between two stocks, we use a partial correlation measurement to determine the underlying relationships between financial assets. Building on previous work, we present a statistically robust approach to extract the underlying relationships between stocks from four different financial markets: the United States, the United Kingdom, Japan, and India. This methodology provides new insights into financial market dynamics and uncovers implicit influences in play between stocks. To demonstrate the capabilities of this methodology, we (i) quantify the influence of different companies and, by studying market similarity across time, present new insights into market structure and market stability, and (ii) we present a practical application, which provides information on the how a company is influenced by different economic sectors, and how the sectors interact with each other. These examples demonstrate the effectiveness of this methodology in uncovering information valuable for a range of individuals, including not only investors and traders but also regulators and policy makers.

\begin{keywords}
Financial markets; Partial correlations; Influence; Risk
\end{keywords}

\begin{classcode}G10, C10, D40\end{classcode}

\end{abstract}

\newpage

\section{Introduction}

Understanding the complex nature of financial markets remains a great challenge, especially in light of the most recent crisis of 2008. Recent studies have investigated large data sets of financial markets, and have analyzed and modeled the static and dynamic behavior of this very complex system \citep{fama1965behavior,lo1997econometrics,lo1990econometric,brock2009more, cont2000herd, eisler2006scaling, lux1999scaling,bouchaud2003theory, voit2005statistical, sinha2010econophysics, abergel2011econophysics, takayasu2006practical,sornette2004stock}, suggesting that financial markets exhibit systemic shifts and display non-equilibrium properties.

One prominent feature in financial markets is the presence of an observed  correlation (positive or negative) between the price movements of different financial assets. The presence of a high degree of cross-correlation between the synchronous time evolution of a set of equity returns is a well known empirical fact \citep{markowitz1952portfolio,elton2009modern,campbell1997econometrics}. 
The Pearson correlation coefficient \citep{pearson1895contributions} provides information about the similarity in the price change behavior of a given pair of stocks. Much effort has been devoted to extracting meaningful information from the observed correlations in order to gain insights into the underlying structure and dynamics of financial markets \citep{embrechts2002correlation,morck2000information,campbell2008increasing,Krishan2009,aste2010correlation,campbell2008increasing,cizeau2001correlation,laloux2000random,laloux1999noise,plerou1999universal,podobnik2009cross,pollet2010average,tumminello2010correlation,huang2013cascading,forbes2002no}.

A large body of work has dealt with the systemic risks introduced into a financial system when there is a co-movement of financial assets. To understand how risks propagate through the entire system, many studies have focused on understanding the synchronization in financial markets that is especially pronounced during periods of crisis \citep{haldane2011systemic,bisias2012survey}. Recent advancements include the CoVaR methodology \citep{adrian2011covar}, and and Granger causality analysis \citep{granger1969investigating,billio2012econometric}. These measures focus on the relationship of one variable on a second variable, for a given time period. Finally, much work has been focused on the issue of conditional correlation \citep{engle2002dynamic} and event conditional correlation \citep{maugis2014event}, and its applications in financial markets. However, a missing dimension of these methodologies is the investigation of many-body interaction between financial assets.

Despite the meaningful information provided by investigating the correlation coefficient, it lacks the capacity to provide information about whether a different stock(s) eventually controls the observed relationship between other stocks. To overcome this issue we introduce the use of the partial correlation coefficient \citep{baba2004partial}, and its applications.  

A partial (or residual) correlation measures how much a a given variable, say $j$, affects the correlations between another pair of variables, say $i$ and $k$. Thus, in this $(i,k)$ pair, the partial correlation value indicates the correlation remaining between $i$ and $k$ after the correlation between $i$ and $j$ and between $k$ and $j$ have been subtracted. Defined in this way, the difference between the correlations and the partial correlations provides a measure of the {\it influence} of variable $j$ on the correlation $(i,k)$. Therefore, we define the influence of variable $j$ on variable $i$, or the dependency of variable $i$ on variable $j$, as $D(i,j)$, to be the sum of the influence of variable $j$ on the correlations of variable $i$ with all other variables. This methodology has originally been introduced for the study of financial data \citep{kenett2010dominating,Kenett2012b,Kenett2012c,maugis2014event}, it has been extended and applied to other systems, such as the immune system \citep{Madi2011b}, and semantic networks \citep{Kenett2011c}. Causality, and more specifically the nature of the correlation relationships between different stocks, is a critical issue to unveil. The main goal of our study is to understand the underlying mechanisms of influence that are present in financial markets.

Previous work has focused on how variable $j$ affects variable $i$, by averaging over all $(i,k)$ pairs, thus quantifying how variable $j$ affects the average correlation of $i$ with all other variables. Although this has provided important information that has been both investigated and statistically validated, our goal here is to present a more general and robust method to statistically pick the meaningful links without first averaging over all pairs. Unlike the previous work in which the average influence of $j$ on the correlation of $i$ with all others was calculated, and then statistically validated, here we first filter for validated links, and then average the influence. In order to achieve this we expand  the original methodology and use statistical validation methods to filter the significant links. This statistically validated selection process reveals significant influence relationships between different financial assets. This new methodology allows us to quantify the influence different assets (e.g., economic sectors, other markets, or macro-economic factors) have on a given asset. The information generated by our methodology is applicable to such areas as risk management, portfolio optimization, and financial contagion, and is valuable to both policy makers and practitioners.

The rest of this paper is organized as follows: In Section 2 we introduce the partial correlation approach to quantify influence between financial assets. We present the new extensions of the methodology, which allows the selection of statistically significant influence links between different assets. We further discuss the empirical data analyzed in this study. In Section's 3 and 4 we present two possible applications of the methodology. In Section 3 we focus on how the methodology provides new insights into market structure and its stability across time, while in Section 4 we present a practical application, which provides information on the how a company is influenced by different economic sectors, and how the sectors interact with each other. Finally, in Section 5 we discuss our results and provide additional insights into possible applications of this methodology.

\section{Quantifying underlying relationships between financial assets}

The aim of this paper is to present a new methodology that sheds new light on the underlying relationships between financial assets. Building on previous work (\cite{kenett2010dominating}), we present a robust and statistically significant approach to extracting the hidden underlying relationships. As such, the presented methodology provides new insights into the underlying mechanisms of financial markets.

\subsection{Data}

For the analysis reported in this paper we use daily adjusted closing stock price time series from four different markets, data provided by the Thomson Reuters Datastream. The markets investigated are the U.S., the U.K., Japan, and India, (see Table 1 for details, also \cite{Kenett2012c}). We only consider stocks that are active from January 2000 until December 2010. Volume data was used to identify and eliminate illiquid stocks from the sample. Table 1 presents the number of stocks remaining after filtering out the stocks that had no price movement for more than 6 percent of  the 2700 trading days. 

\begin{table}[!h]
\begin{center}
\begin{minipage}{80mm}
\tbl{Summary of data sample}
{\begin{tabular}{ c | c | c | c | c }
\hline
  Market & Stocks used & Index used & \# before & \# filtered \\\hline
  U.S.	& S\&P 500	& S\&P 500	& 500	& 403\\
  U.K.	& FTSE 350	& FTSE 350	& 356	& 116\\
  Japan	& Nikkei 500	& Nikkei 500	& 500	& 315\\
  India	& BSE 200	& BSE 100	& 193	& 126\\
       \hline
\end{tabular}}
\end{minipage}
\end{center}
\label{table_data}
\end{table}

\subsection{Stock raw and partial correlation}
To study the similarity between stock price changes, we calculate the time series of the daily log return, given by
\begin{equation}
r_i(t)=log[P_i(t)/P_i(t-1)],
\end{equation}
where $P_i(t)$ is the daily adjusted closing price of stock $i$ at day $t$. The stock raw correlations are calculated using the Pearson correlation coefficient~\citep{pearson1895contributions}

\begin{equation}
\rho(i,j)=\frac{\left\langle r(i)-\left\langle r(i) \right\rangle \right\rangle ~\cdot~ \left\langle r(j)-\left\langle r(j) \right\rangle \right\rangle}{\sigma(i) ~\cdot~\sigma(j)},
\end{equation}
where $\left\langle \right\rangle$ represents average over all days, and $\sigma(i)$ denotes the standard deviation.

However, in some cases, a strong correlation not necessarily means strong direct relation between two variables. For example, two stocks in the same market can be influenced by common macroeconomic factors and investor psychological factors. To study the direct correlation of the performance of these two stocks, we need to remove the common driving factors, which are represented by the market index. Partial correlation quantifies the correlation between two variables, e.g. stocks returns, when conditioned on one or several other variables~\citep{baba2004partial,shapira2009index, kenett2010dominating}. Specifically, let $X$, $Y$ be two stock return time series and $M$ be the index. The partial correlation, $\rho(X, Y: M)$, between variables $X$ and $Y$ conditioned on variable $M$ is the Pearson correlation coefficient between the residuals of $X$ and $Y$ that are uncorrelated with $M$. To obtain these residuals of $X$ and $Y$, they are both regressed on $M$. The partial correlation coefficient can be expressed in terms of the Pearson correlation coefficients as

\begin{equation}
\rho(X,Y:M)\equiv \frac{\rho(X,Y)-\rho(X,M)\rho(Y,M)}{\sqrt{\left[1-\rho ^2(X,M)\right]\left[1-\rho ^2(Y,M)\right]}}.
\label{eq_part_corr}
\end{equation} 

\begin{figure}[h]
\centering
    \subfigure[]{
    \includegraphics[width=0.45\textwidth ]
	{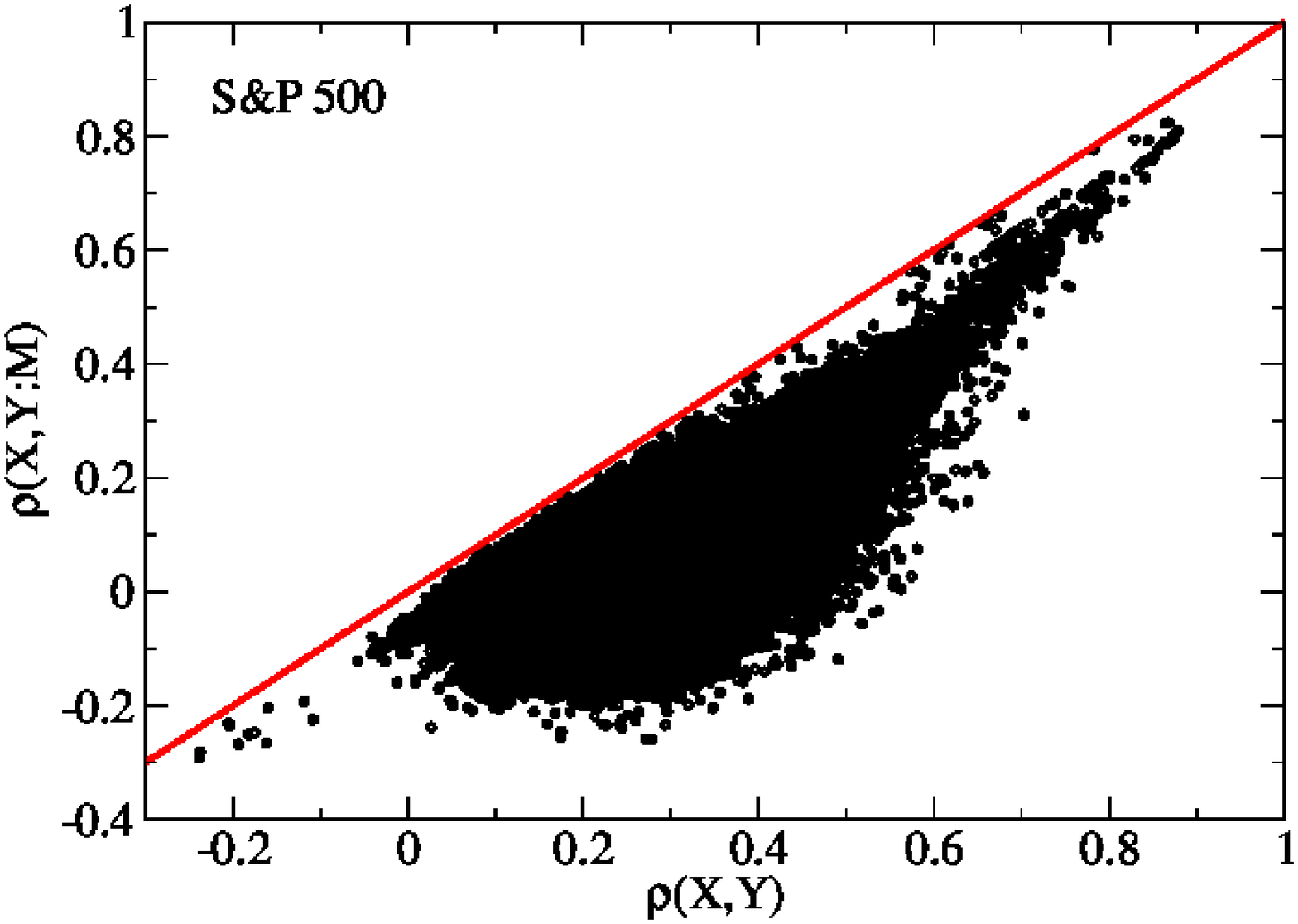}
	}
    \subfigure[]{
    \includegraphics[width=0.45\textwidth ]
	{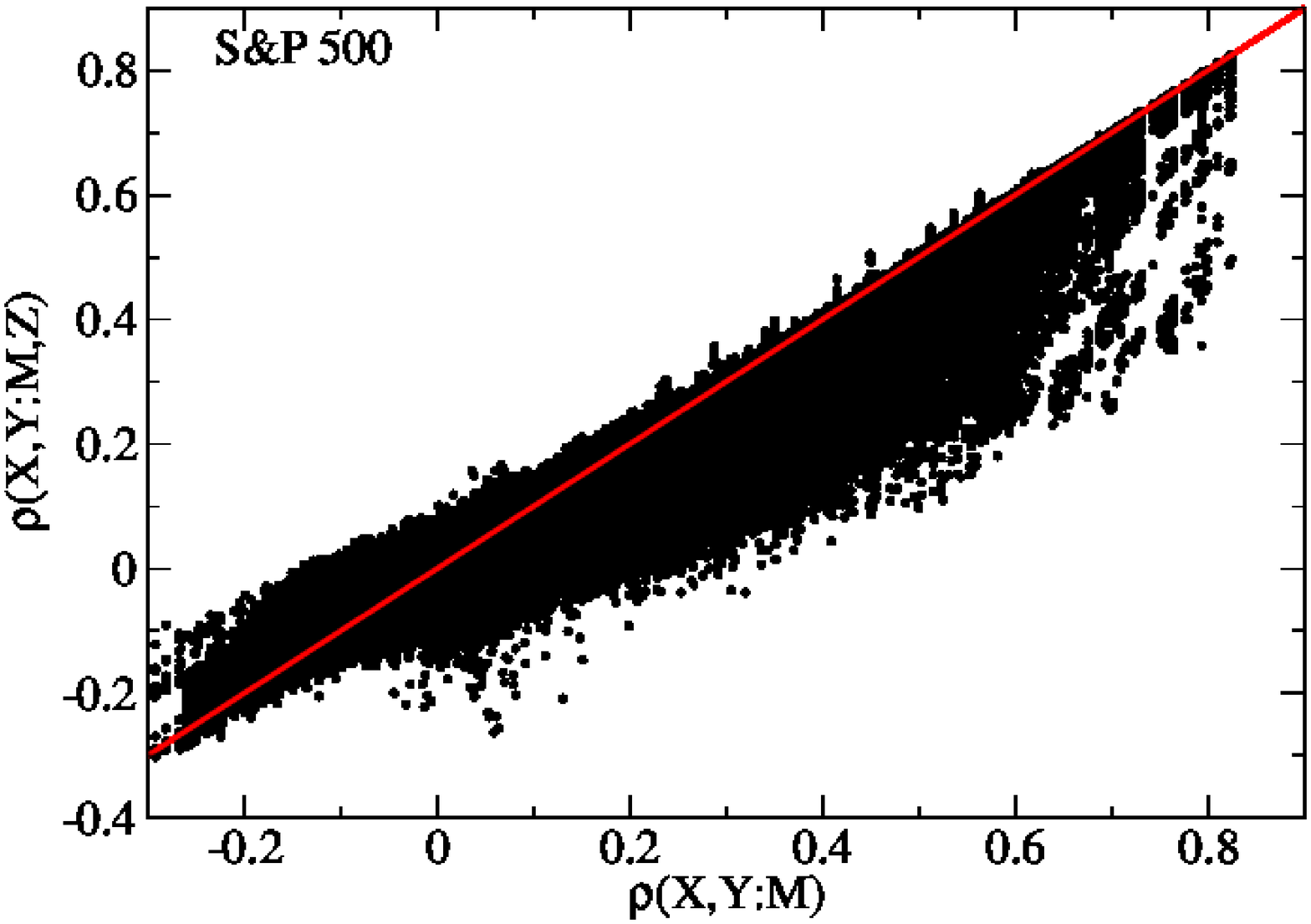}
	}
	\caption{(a) Scatter plot of correlation {\it v.s.} partial correlation conditioned on index. It is possible to observe that all points are below the diagonal straight line, which means the influence from the index to the correlation between any pair of stocks is always positive. (b) Scatter plot of partial correlation conditioned on index {\it v.s} partial correlation conditioned on both index and a third stock. It is possible to observe that the points distribute at both sides of the diagonal line, meaning a significant fraction of $d(X,Y:Z)$ is negative (mainly for low $\rho(X, Y:M)$ values). In both figures, $X, Y, Z$ represent return time series of stocks, $M$ represents return time series of S\&P 500 index. The red curve is the diagonal straight line.}
	\label{fig_corr_scatter_plot}
\end{figure}

In figure~\ref{fig_corr_scatter_plot} (a), we plot the correlation and partial correlation (using the index as the conditioning variable) between stocks that belong to the S\&P 500 index. The figure shows that all points are below the diagonal straight line, which means the influence from the index to the correlation between any pair of stocks is always positive.
Furthermore, when two stocks $X$ and $Y$ both have business relation with a third common stock $Z$, their prices can be both affected by the performance of the third stock, thus showing similar price movements even after removing the index. By removing the influence from the third company, we can see the importance of the role that the third stock acts in the correlation of two stocks. Partial correlation coefficient between $X$ and $Y$ conditioned on both $M$ and $Z$ is

\begin{equation}
\rho(X,Y:M,Z)\equiv \frac{\rho(X,Y:M)-\rho(X,Z:M)\rho(Y,Z:M)}{\sqrt{\left[1-\rho ^2(X,Z:M)\right]\left[1-\rho ^2(Y,Z:M)\right]}}.
\label{eq_part_corr2}
\end{equation}

In order to quantify the influence of stock $Z$ on the pair of $X$ and $Y$, we focus on the {\it Influence} quantity
\begin{equation}
d(X,Y:Z)\equiv \rho(X,Y:M)-\rho(X,Y:M,Z).
\label{eq_influence}
\end{equation}
This quantity is large when a significant fraction of the partial correlation $\rho(X,Y:M)$ can be explained in terms of $Z$. In previous research, Kenett {\it et al.} defined this quantity by $d^*(X,Y:Z)\equiv \rho(X,Y)-\rho(X,Y:Z)$, which holds for general cases  \citep{kenett2010dominating}. However, for the stock market case specifically, the fraction of $\rho(X,Y)$ that can be explained by $Z$ contains two parts, index influence and stocks $Z$ influence, because stock $Z$ contains information of the index. Usually, the influence from the index is prevailing and overwhelms the influence from an individual stocks. For example, when $X$ and $Y$ are competitor and cooperator respectively to stock $Z$, performances of $X$ and $Y$ should have negative correlation because of $Z$. In this case, the influence from $Z$ to the correlation between stocks $X$ and $Y$ should be negative. However, because of the dominant correlation between these two stocks and the index, the $d^*(X,Y:Z)$ is still positive. Thus, we suggest to remove the influence of the market before studying the influence of a stock $Z$ on a pair of stocks. In the scatter plot of the partial correlation conditioned on index v.s. the partial correlation conditioned on both index and an individual stock (figure~\ref{fig_corr_scatter_plot} (b)), the points distribute at both sides of the diagonal line, meaning a significant fraction of $d(X,Y:Z)$ is negative.

The average influence $d(X:Z)$ of stock $Z$ on the correlations between stock $X$ and all the other stocks in the system is defined as

\begin{equation}
d(X:Z)\equiv \left\langle d(X,Y:Z) \right\rangle_{Y\neq X}.
\label{eq_influence_stock_to_stock}
\end{equation}
It is important to note that $d(X:Z)$ approximates the net influence from stock $Z$ to stock $X$, excluding the influence from the index.

\subsection{Test of statistical significance}
In a system of size $N$, there exists $N(N-1)(N-2)/2$ partial correlation interactions, $d(X,Y:Z)$, when all information is considered. To simplify the description of the system, the non-trivial interactions with certain significance level are selected. To decide the significance of partial correlation, we provide two methods: 1) Fisher's transformation based approach; and 2) empirical based approach.

\subsubsection{Fisher transformation statistical significance test}
We first introduce the Fisher's transformation method. According to ref.~\citep{fisher1915frequency}, when $X$ and $Y$ follow a bivariate normal distribution and $X(t)$, $Y(t)$ pairs to form the correlation are independent for $t=1...n$, a transformation of the Pearson correlation 
\begin{equation}
z(\rho) = {1 \over 2}\ln\left({1+\rho \over 1-\rho}\right) = \operatorname{artanh}(\rho)
\label{eq_Fisher_Trans}
\end{equation}
approximately follows normal distribution $\mathcal{N}({1 \over 2}\ln\left({1+r \over 1-r}\right), {1 \over \sqrt{N-3}})$, where $r$ is the population correlation coefficient and $N$ is the sample size. The Fisher transformation holds when $\rho$ is not too large and $N$ is not too small. 
Furthermore, the Fisher's z-transform of the partial correlation coefficients approximately follows~\citep{fisher1924distribution}

\begin{equation}
\left\lbrace
\begin{split}
z\left( \rho(X,Y:M) \right) &\sim \mathcal{N}({1 \over 2}\ln\left({1+r(X,Y:M) \over 1-r(X,Y:M)}\right), {1 \over \sqrt{N-3}})\\
z\left(\rho(X,Y:M,Z)\right) &\sim \mathcal{N}({1 \over 2}\ln\left({1+r(X,Y:M,Z) \over 1-r(X,Y:M,Z)}\right), {1 \over \sqrt{N-3}})
\end{split}
\right.
\end{equation}
which leads to 

\begin{align}
z(\rho(X,Y:M))- & z(\rho(X,Y:M,Z)) \sim \nonumber \\
 & \mathcal{N}\left({1 \over 2}\ln\left({1+r(X,Y:M) \over 1-r(X,Y:M)}\right)-  {1 \over 2}\ln\left({1+r(X,Y:M,Z) \over 1-r(X,Y:M,Z)}\right), {\sqrt{{2 \over N-3}}}\right).
 \label{zequation}
\end{align}
When $\rho(X,Y:M)$ is significantly different from $\rho(X,Y:M,Z)$, $d(X,Y:Z)$ is significantly different from zero. Thus the Student's t-test is used to determine if $\rho(X,Y:M)$ and $\rho(X,Y:M,Z)$ are different. Since $N$ is large, the degree of freedom is also large, where the t-test is approximately the same as the Z-test \citep{chou1975statistical}. In the next section, we propose a complementary empirical statistical significance test. The empirical approach overlaps with the Fisher transformation approach, and is faster and simpler to evaluate. For real finite size data the two are interchangeable, and the empirical approach is simpler to implement. Thus, after introducing the empirical approach below, we will make use of it throughout the remainder of the paper.

\subsubsection{Empirical statistical significance test}
\begin{figure}[h]
\includegraphics[width=0.9\textwidth]{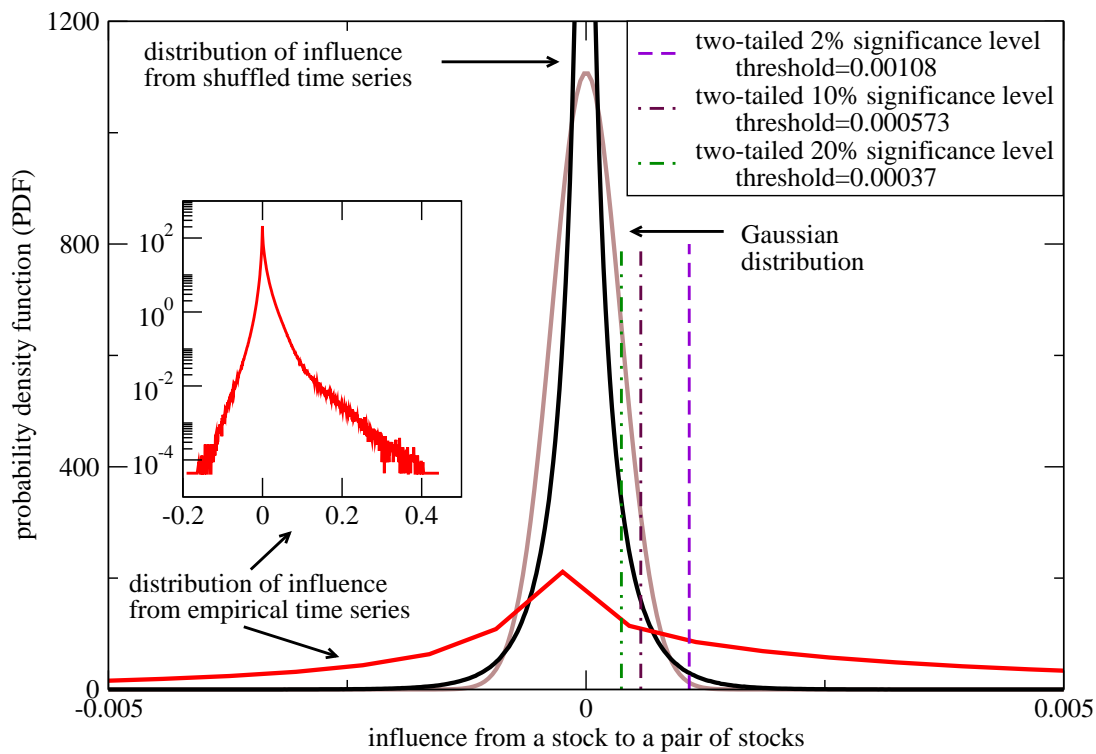}
\caption{Empirical statistical significance test of influence between financial assets. We present the distribution of influence from one stock to pair of stocks, for the case study of 403 S\&P500 companies, for a period of 11 years. The black curve represents the distribution of the influences, $\hat d(X,Y:Z)$, calculated from the shuffled time series. The dashed lines denotes the positions of one-tailed $1\%$, $5\%$ and $10\%$ significance levels or two-tailed $2\%$, $10\%$ and $20\%$ significance levels. The distribution of $\hat d(X,Y:Z)$ has an average value $9.9 \cdot 10^{-8}$, standard deviation $\sigma = 3.6 \cdot 10^{-4}$, skewness $9.5 \cdot 10^{-7}$ and kurtosis $9.83$. The brown curve represents the Gaussian distribution with the same average value and standard deviation as the black curve. The red curve is the distribution of influences, $d(X,Y:Z)$, calculated from the empirical stock return time series. We suggest to make use of the two-tailed test, due to the fact that the significant negative influence values are also important. For the two-tailed test, significance level of $2\%$ corresponds to $z>1.6449$; confidence level of $10\%$ corresponds to $z>1.2816$; and confidence level of $20\%$ corresponds to $z>0.8416$. The inset shows the full empirical distribution.}
\label{fig_dist_influence_sf}
\end{figure}

Next, we introduce the empirical time series shuffling statistical significance method. For each time series that we study, we shuffle the sequences of the returns, which destroys any correlations among these time series. The influences $\hat d(X,Y:Z)$ calculated from these time series should also be not different than zero. We shuffle the return time series of each stock by randomly rearranging the sequences of the returns. The shuffling process destroys correlations between each pairs of stocks returns, and between stock returns and benchmark return, i.e. $\rho (X, Y)$ and $\rho (X, M)$ should be zero, which leads to $d(X, Y : Z)$ equal to zero. By plotting the distribution of $\hat d(X,Y:Z)$, we can find the thresholds of different significant levels for $d(X,Y:Z)$. In fig.~\ref{fig_dist_influence_sf}, we shuffle the return time series of 403 S\&P500 stocks, and the index, and plot the distribution of $\hat d(X,Y:Z)$ (solid black curve).  As a comparison, we plot the Gaussian distribution with same average and standard deviation value in brown color. From the comparison, it is possible to observe that the empirical distribution of of $\hat d(X,Y:Z)$ has fat tails, and significantly deviates from the case of a gaussian distribution observed for random or shuffled data. The dashed lines represent the positions of one-tailed $1\%$, $5\%$ and $10\%$ significance levels or two-tailed $2\%$, $10\%$ and $20\%$ significance levels. We suggest to make use of the two-tailed test, due to the fact that the significant negative influence is also important. The red curve in fig.~\ref{fig_dist_influence_sf} presents the distribution of empirical $d(X,Y:Z)$. For the case of the S\&P500 companies example presented in fig.~\ref{fig_dist_influence_sf}, significance level of $2\%$ corresponds to a $z-value$ (Equation~\ref{zequation}) $z>1.6449$; confidence level of $10\%$ corresponds to $z>1.2816$; and confidence level of $20\%$ corresponds to $z>0.8416$ (see Table 2).

\begin{table}[!h]
\begin{center}
\begin{minipage}{80mm}
\tbl{Summary of two-tail significance thresholds}
{\begin{tabular}{ | c | c | }
\hline
  Significance level & Threshold \\\hline
  1\%	& 0.00152\\
  2\%	& 0.00108\\
  5\%	& 0.00081\\
  10\%	& 0.00057\\
  20\% & 0.00037\\
       \hline
\end{tabular}}
\end{minipage}
\end{center}
\label{tableth}
\end{table}

As discussed above, while the two methods are interchangeable in respect to the resulting significance levels, there are additional benefits to using the empirical approach. The empirical shuffling method conserves the distribution of returns without requiring ad-hoc assumptions on the underlying distribution of the data. Furthermore, the empirical approach can also be expanded to meet stricter requirements. For example, if the short term time series structure is required to be conserved, the whole time series can be divided into segments with a given length and we can than shuffle the segments without changing the sequence of time series within the segments.

The empirical statistical significance test allows a selection of significant influence relationships between the investigated financial assets. In previous work (\cite{kenett2010dominating}), this was achieved by using different network based approaches, which than further allowed to investigate the nature of these relationships. Below we propose two new applications of this methodology, using the empirically statistically significant values of $d(X,Y:Z)$. The significance level used through out the paper is the $z>1.6449$, corresponding to two-tailed 2\%.

\section{Market structure and its stability}
High correlation between two stocks at a given time do not necessarily guarantee high correlation in the future, because the behavior of stocks in financial markets evolve. In certain markets, companies change their strategies faster than in the other markets, which can be uncovered by the partial correlation analysis of the behavior of their stocks. If the companies tend to keep their past strategies, then the level of partial correlation between two companies' stocks tends to be stable. While in markets where companies switch their strategies more quickly, two companies which had similar behavior in the previous year might have quite different behavior in the next year. In such markets, partial correlation between stocks should be more volatile.

We apply the partial correlation influence analysis to study the stability of the market structure. Specifically, we define the average influence $d(X)$ of stock $X$ on all the other stocks in the market as
\begin{equation}
d(X)\equiv\left\langle d(X:Z)\right\rangle,
\end{equation}
where $\left\langle \right\rangle$ is the average over all $Z$ stocks. We rank the stocks by their $d(X)$ values, which we consider as a representation of the structure of the market. By dividing the 11 year period into 44 quarterly periods, we can compare similarity of the market structures (ranking of stocks) in different years. Kendall $\tau$ rank correlation coefficient~\citep{kendall1938new} is applied to measure the similarity of the orderings for different periods. Let $(x_1(t), x_1(t')), (x_2(t), x_2(t')), ..., (x_n(t), x_n(t'))$ be a set of rankings of the variables $X$ for different periods $t$ and $t'$ respectively. Any pair of observations $(x_i(t), x_i(t'))$ and $(x_j(t), x_j(t'))$ are said to be concordant if both $x_i(t) > x_j(t)$ and $x_i(t') > x_j(t')$ or if both $x_i(t) < x_j(t)$ and $x_i(t') < x_j(t')$. Otherwise, they are said to be discordant.
The Kendall $\tau$ coefficient is defined as
\begin{equation}
\tau = \frac{\mbox{number of concordant pairs} - \mbox{number of discordant pairs}}{\frac{1}{2}n(n-1)},
\end{equation}
where if two rankings are the same, $\tau$ is one, if two rankings are independent, $\tau$ is zero, and if two rankings are discordant, $\tau$ equals minus one.

\begin{figure}[!h]
\includegraphics[width=0.9\textwidth]{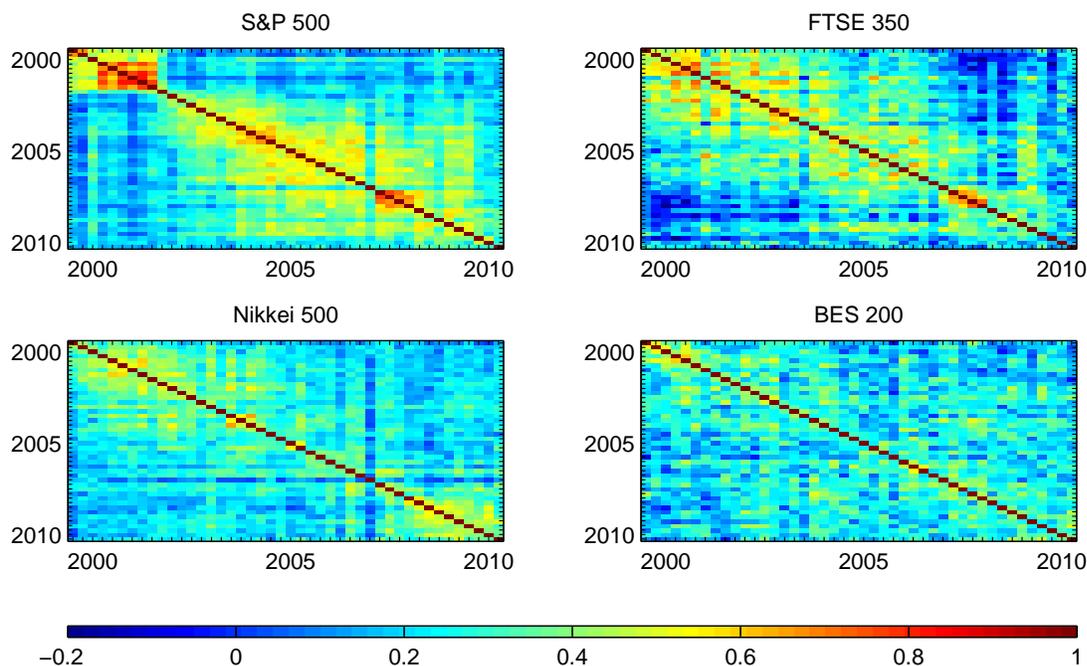}
\caption{Kendall $\tau$ correlation coefficient of stock rankings in different quarters. The total investigated period of 11 years is broken down into 44 quarters. For each quarter, we calculate the average influence of each stock, and rank the stocks according to their influence. We then use the Kendal $\tau$ correlation to quantify the similarity between every possible quarter pair, for the four investigated markets. The color code is used to present the strength of the correlation, ranging from blue for negative rank correlation to red for positive rank correlation.}
\label{fig_kendall_coefficient}
\end{figure}

In figure~\ref{fig_kendall_coefficient}, we present the Kendall $\tau$ coefficient for each different quarter pairs for the four investigated markets. Generally speaking, each market shows that the longer the time interval the smaller the rank correlation coefficient, meaning lower similarity between the market structures for the two quarters which are compared. Comparing the rank correlations for the four markets, we find that S\&P 500, FTSE 350 and Nikkei 500 stocks show strong market stability patterns, while Indian BES 200 stocks almost do not demonstrate any stable patterns. This can be understood by considering that developed markets tend to keep their market structure longer than fast developing markets. Furthermore, it is possible to observe that for the US market there were structural changes in the market following the \lq\lq dot com\rq\rq crisis of 2000 and the \lq\lq credit crunch\rq\rq crisis of 2008. These can be identified in figure~\ref{fig_kendall_coefficient} by the red rectangle in the upper left corner for the former (Q4 of 2000 till Q4 of 2001), and the red rectangle in the bottom right corner for the latter (Q4 of 2007 till Q4 of 2008). These rectangles present a strong similarity in the structure during the two crises, followed by consecutive quarters with low values of thee rank correlation, representing the change in structure. Studying the other markets, it is also possible to observe the structural changes resulting from the 2008 financial crisis in the UK, but not in the structure of Japan or India.

\begin{figure}[!h]
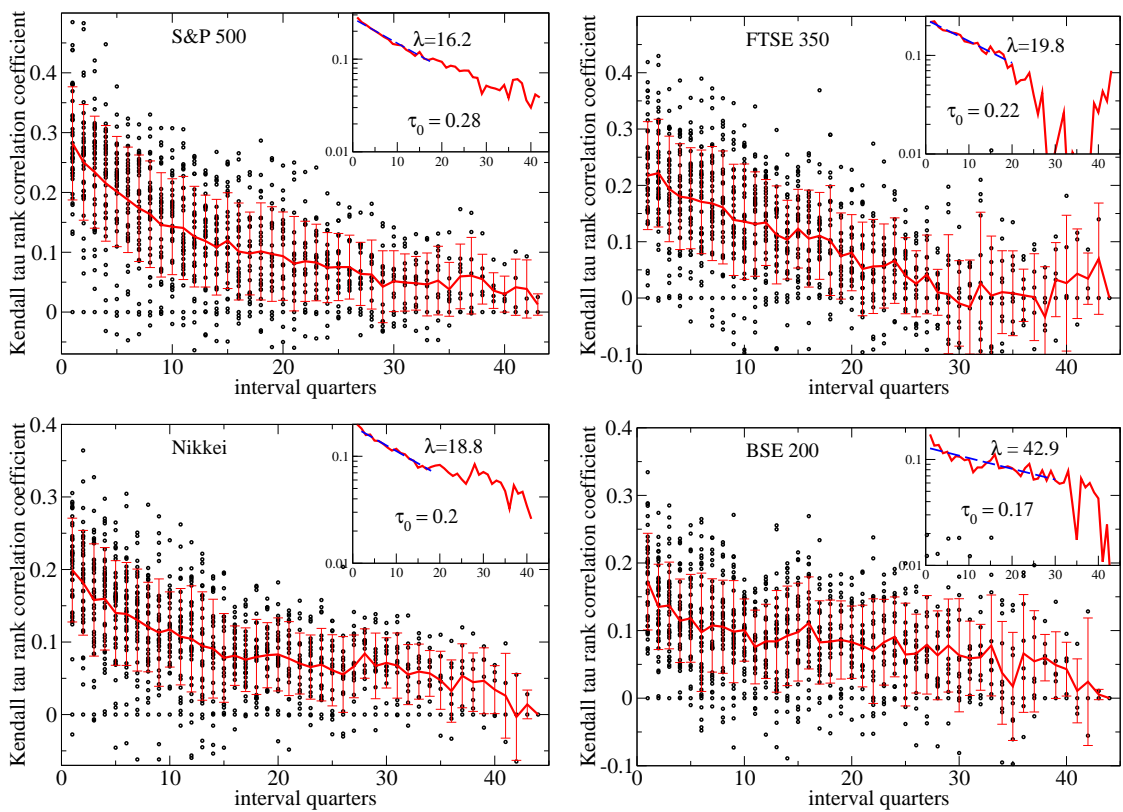

\centering
    \subfigure{
    \includegraphics[width=0.45\textwidth ]
	{plot_quarter_kendallDist_sp.eps}
	}
    \subfigure{
    \includegraphics[width=0.45\textwidth ]
	{plot_quarter_kendallDist_ftse350.eps}
	}
	\subfigure{
    \includegraphics[width=0.45\textwidth ]
	{plot_quarter_kendallDist_nikkei.eps}
	}
	\subfigure{
    \includegraphics[width=0.45\textwidth ]
	{plot_quarter_kendallDist_india.eps}
	}
\caption{Kendall tau correlation coefficient between rankings {\it v.s.} time interval between rankings. By averaging the correlation coefficients for each time interval, we can study how correlation coefficients decay as time evolves. We find the decay of the $\tau$ rank correlation coefficients follow an approximate exponential process, $\tau = \tau _0 e^{-t/\lambda}$. Parameter $\tau _0$ describes the consistency of the rankings between two and  consecutive quarters. The larger $\tau _0$ is, the more consistent two consecutive ranking are. The $\lambda$ describes the speed that the correlation coefficient decays as time evolves. Larger $\lambda$ means longer decay period and smaller decay speed. These two parameters together describes the stability of the markets. For the investigated markets, we obtain the following values: US - $\tau_0=0.28, \lambda=16.2$; UK - $\tau_0=0.22, \lambda=19.8$; Japan - $\tau_0=0.2, \lambda=18.8$; and India - $\tau_0=0.17, \lambda=42.9$.}
\label{fig_kendall_coefficient2}
\end{figure}

To further quantitatively study the market stability, we plot the correlation coefficient of two rankings against the time interval of these two rankings (fig.~\ref{fig_kendall_coefficient2}). By averaging the correlation coefficients for each time interval, we can study how correlation coefficients decay as time evolves. We find the decay of the $\tau$ rank correlation coefficients follow an approximate exponential process, $\tau = \tau _0 e^{-t/\lambda}$, as shown in the insets in fig.~\ref{fig_kendall_coefficient2}. Parameter $\tau _0$ describes the consistency of the rankings between two consecutive quarters. The larger $\tau _0$ is, the more consistent two consecutive ranking are. The $\lambda$ parameter describes the characteristic time after which the correlation coefficient decays. Larger $\lambda$ values mean longer persistence period, and thus describe the change in influence ranking across time. These two parameters together describes the stability of the markets. For the investigated markets, we obtain the following values: US - $\tau_0=0.28, \lambda=16.2$; UK - $\tau_0=0.22, \lambda=19.8$; Japan - $\tau_0=0.2, \lambda=18.8$; and India - $\tau_0=0.17, \lambda=42.9$. As can also be observed in figure~\ref{fig_kendall_coefficient}, $\tau_0$ has the largest value for the S\&P500 case, and smalls value for the BES200 case; however, the persistence in India is largest (as represented by the values of $\lambda$).  We observe that the results for the Indian market differ from the other three markets. This is possibly related to the differences observed between developing and developed markets.

Put together, these analyses provide new insights into the dynamics of financial markets. Using the $\tau_0$ and $\lambda$ parameters, can help in monitoring structural changes in the market, and their persistence. Thus, this methodology presents a unique tool for regulators and policy makers to monitor the stability and robustness of financial markets. 

\section{Quantifying the influence of economic sectors}
As our society becomes more and more integrated, production activities from different industries depend upon and influence each other. Categorizing a company into only one industrial sector can not reflect its whole performance and associated risk. Many listed companies in the stock market belong to conglomerates, conducting their business in different industry sectors, hence these companies' performance will naturally be influenced by multiple industries. Even if a company only conducts its business in one sector, its performance can still be influence by other sectors because of the division of labor in modern society. For example, Alcoa Inc. as the world's third largest producer of aluminum is listed in the materials sector in NYSE. However, the production of Alcoa Inc. requires dedicated supply of energy, e.g. Alcoa accounts for 15\% of State of Victoria's annual electricity consumption in Australia. Thus their performance is also heavily influenced by and contributes to the performance of the energy sector. In this section, we present an application of the partial correlation methodology to study the multiple-sector influence on stocks. We use the sector classification from the Global Industry Classification Standard (GICS).

\begin{figure}[!h]
\centering
\includegraphics[width=0.8\textwidth]{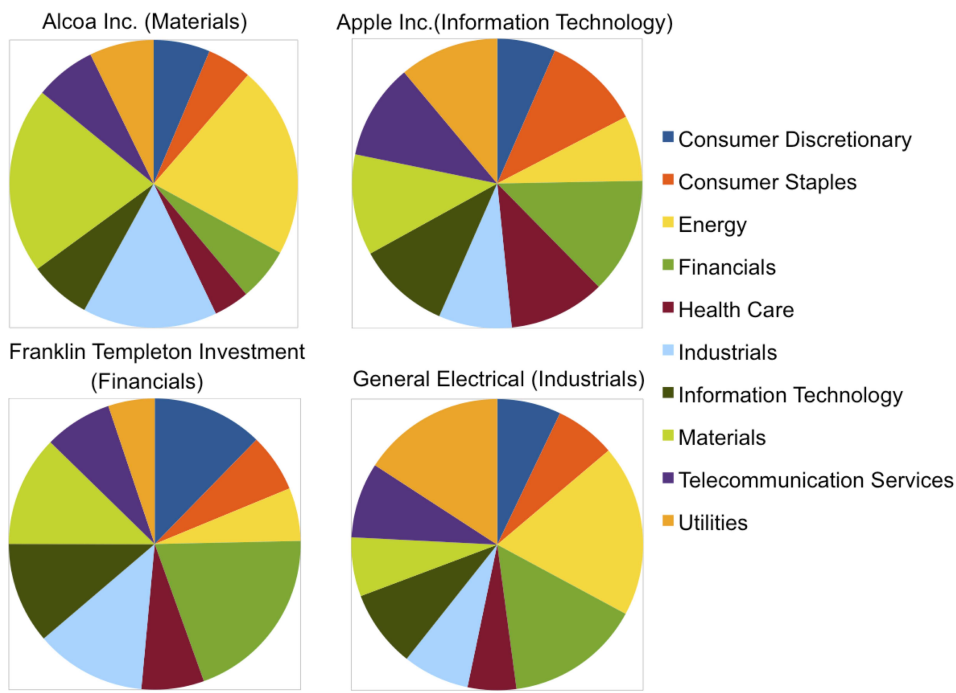}
\caption{Fraction of influence from each sector to example stocks, Alcoa Inc., Apple Inc., Franklin Templeton Investment and General Electrical. we present an example of four typical stocks to show the pie picture of $\beta ^S_X$. We can see from the figure that in the case of Alcoa Inc., we observe significant influence from the energy, materials and industrials sector. In the case of Franklin Templeton Investments, we find that the largest influence is from the financials sector. In the case of GE, we find that the main influence stems from the materials, utilities, and financial sector. Finally, studying the example of Apple, we find that there is a more homogeneous division of the influence between the different sectors. This could possibly indicate that out of these 4 companies, Apple is the most diverse, as its business is affected by different sectors of the economy. }
\label{fig_components}
\end{figure}

To study the influence on a stock $X$ from different sectors, we first calculate the influence $d(X:Z)$ (eq.~\ref{eq_influence_stock_to_stock}) from all other stocks $Z$. The analysis in this section is performed for the entire investigated time period. Next we calculate the average influence by sector, in which we use the sector categorization information of other stocks, as follows
\begin{equation}
d^S_X = \frac{1}{N_S} \sum_{Z_S=1}^{N_S} d(X:Z_S),
\end{equation}
where $X$ represents the investigated stock, $S$ represents a given sector, $N_S$ is the number of stocks in sector $S$ and $Z_S$ represents the stocks in sector $S$. The average influence $d^S_X$ reflects the level of influence that stock $X$ receives from sector $S$. After we normalize the average influence, we can attribute stock $X$'s performance to sectors' performances with coefficients
\begin{equation}
\beta ^S_X = \frac{d^S_X}{\sum_{S}d^S_X}.  
\end{equation}
In figure~\ref{fig_components}, we present an example of four typical stocks to show the pie picture of $\beta ^S_X$. We can see from the figure that in the case of Alcoa Inc., we observe significant influence from the energy, materials and industrials sector. In the case of Franklin Templeton Investments, we find that the largest influence is from the financials sector. In the case of GE, we find that the main influence stems from the materials, utilities, and financial sector. Finally, studying the example of Apple, we find that there is a more homogeneous division of the influence between the different sectors. This could possibly indicate that out of these 4 companies, Apple is the most diverse in its activities, being influenced  almost uniformly by different sectors of the economy. 

\begin{figure}[H]
\centering
    \includegraphics[width=0.8\textwidth ]
	{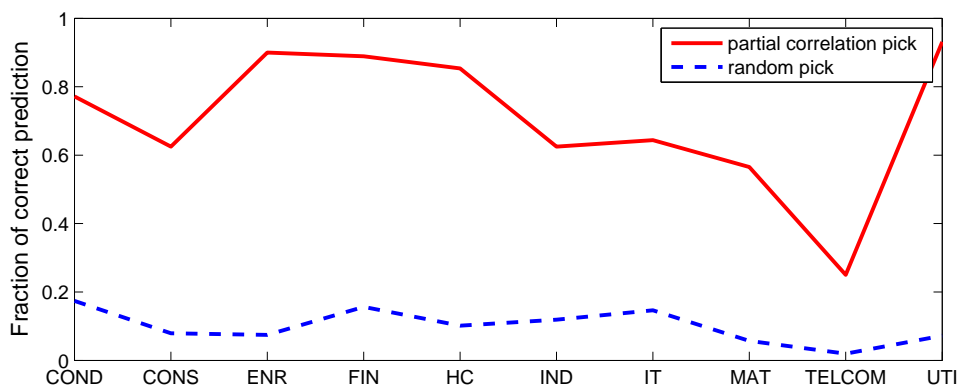}
\caption{Partial correlation test of the extent of sectorial influence. To this end, we first rank all the stocks in S\&P 500 by their fraction of influence ($\beta ^S_X$). We plot the fraction of true prediction of stocks' sector from partial correlation analysis (red solid curve). The blue dashed curve is for the case of random picking strategy, for the purpose of comparison. }
\label{fig_prediction_rate}	
\end{figure}

Finally, we perform a validation test on the the partial correlation analysis results, investigating whether the result of multi-sector influence on stocks is plausible. To this end, we first rank all the stocks in the S\&P500 dataset by their fraction of influence ($\beta ^S_X$) from the financials sector. We then investigate what are the economic sectors influencing these stocks, according to the rank. We find that the top stocks in the ranking according to our partial correlation analysis are dominantly classified into the financials sector. We repeat this analysis for all other economic sectors. Indeed, all other sectors show that our analysis is in agreement with the GICS sector classification. To quantitatively show this agreement, we calculate the correct prediction rate. According to the GICS, we find the total numbers of stocks ($N_S$) in all sectors. From the ranking of stocks according to the influence from a given sector, we select the $N_S$ top stocks. We then calculate the fraction of these top stocks that are classified by GICS into that certain sector as the correct prediction rate. If the partial correlation analysis prediction is in total agreement with the formal classification, then this correct prediction rate should be 1. If the prediction corresponds to the case of random picking, this correct prediction rate should be $\frac{N_S}{N}$, where $N$ is the total number of stocks. In figure~\ref{fig_prediction_rate}(c), we show that the partial correlation analysis of sector keeps a high correct prediction rate for all sectors, except the telecommunications sector, which could be related to the small number of telecommunication stocks that are part of the S\&P500 index. An alternative interpretation to these results is that the financials and energy sectors are both highly cohesive sectors of economic activity. This means that the activity in these economic sectors is mainly contained inside each sector, with little influence of external economic sectors. Other sectors, such as Industrials and Materials, have stronger dependencies on other sectors, and are strongly influenced other sectors' activities. Repeating this analysis for shorter periods of time, or combining the analysis with a moving window approach, could provide meaningful insights into the degree of interdependencies between the different economic sectors over time.

\begin{figure}[H]
\centering
    \includegraphics[width=\textwidth ]
	{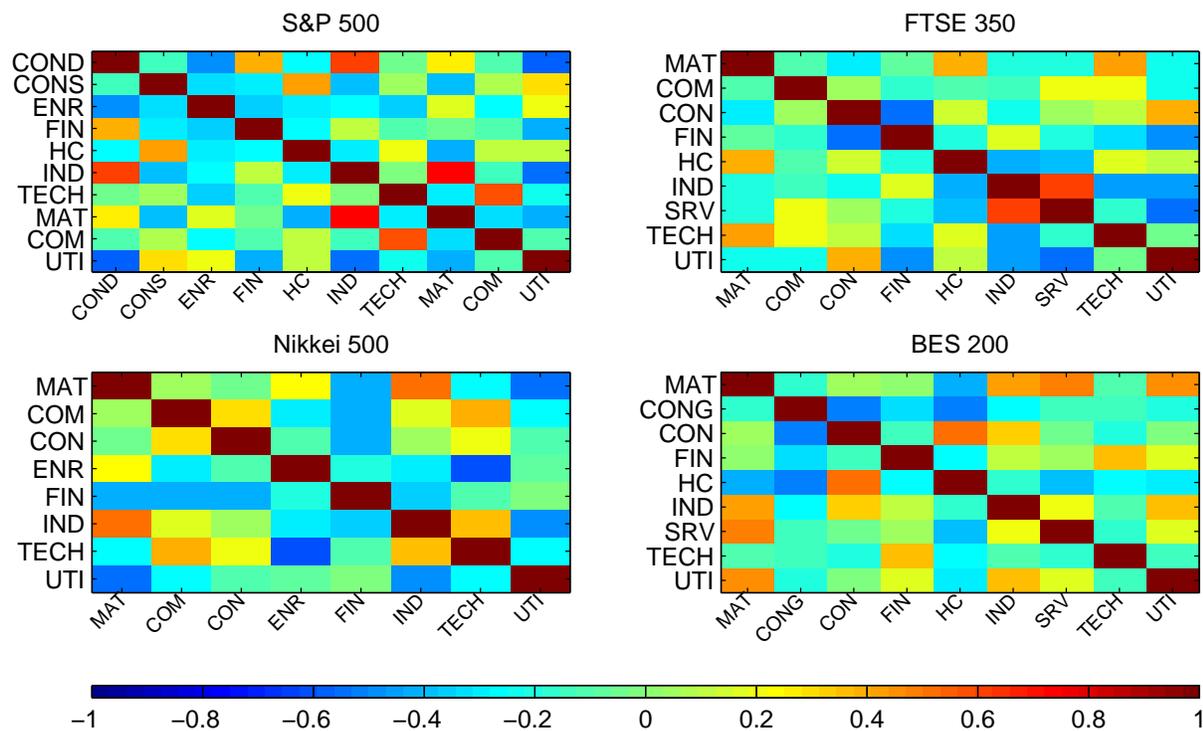}
	\caption{The correlation of influences from two sectors to the same stocks, i.e. $\rho(d^{S_i}, d^{S_j})$, where $d^{S_i}$ represents the vector variable of influence from sector $i$ to all stocks. We represent the correlation using a heat map of closeness between sectors in different markets, using a color code to represent the strength of the correlation, ranging from blue for full anti correlation, to red, for complete correlation.  We observe that for the S\&P 500 data, the pairs of industrials sector and consumer discretionary sector, materials sector and industrials sector and the communications sector and the technology sector are very close to each other, in terms of their influence, and an anti correlation between the utilities sector and the consumer discretionary sector. For the case of the other three investigated markets - the UK FTSE 350 index, the Japanese Nikkei 500 index and the India BES 200 index - we find that all show high correlation between materials sector and industrials sector. }
	\label{sector_closeness_all}
\end{figure}

After studying  the amount of influence that stocks receive from different sectors, we find that some sectors tend to influence the same stocks concurrently. We thus study the Pearson correlation of influences from two sectors to the same stocks, i.e. $\rho(d^{S_i}, d^{S_j})$, where $d^{S_i}$ represents the vector variable of influence from sector $i$ to all stocks. Applying this definition of sector correlation to the S\&P 500 data results in values that are presented in the first panel of figure~\ref{sector_closeness_all}. We find that in the S\&P 500 index, the pairs of industrials sector and consumer discretionary sector, materials sector and industrials sector and the communications sector and the technology sector are very close to each other, in terms of their influence. Whenever a stock is highly influenced by one of these sectors, the other in the pair also tends to be influential to this stock. We also notice some dark blue area, e.g. the correlation between the utilities sector and the consumer discretionary sector, which means when a stocks is highly influence by one of the them, the other tends to be of little influence to the stock. We also study the UK FTSE 350 index, the Japanese Nikkei 500 index and the India BES 200 index. They all commonly show high correlation between materials sector and industrials sector. 

\section{Summary}

This work presents a more general, statistically robust framework of the dependency network methodology introduced by \cite{kenett2010dominating}. Using the dependency network methodology, we apply the partial correlation analysis to uncover dependency and influence relationship between the different companies in the investigated sample. Here we present a new statistically robust approach to filtering the extracted influence relationships, by either using a theoretical or an empirical approach. The influence method introduced in this study is generic and scalable, making it highly accessible to both policy makers and practitioners. 

Here, we present two possible applications of this methodology. First, we study the stability of financial market structure and show that developed markets such as the US, UK, and Japan exhibit higher degree of market stability compared to developing countries such as India. Second, we show that one stock can be influenced by different sectors outside of its primary sector classification.  

While financial analysts are usually specialized in one industry sector, a broader perspective of equity research is required to grasp the insights of stock performance expectations. The presented methodology provides new information on the interaction between different assets, and different economic sectors. Such information is valuable not only for investors and their practitioners, but also for regulators and policy makers. 

\section*{Acknowledgements}
We wish to thank ONR (Grant N00014-09-1-0380, Grant N00014-12-1-0548),
DTRA (Grant HDTRA-1-10-1- 0014, Grant HDTRA-1-09-1-0035), NSF (Grant
CMMI 1125290), the European MULTIPLEX (EU-FET project 317532), CONGAS (Grant
FP7-ICT-2011-8-317672), FET Open Project ÔÔFOCÕÕ 255987 and
ÔÔFOC-INCOÕÕ 297149, and LINC (no. 289447 funded by the ECÕs Marie-
Curie ITN program) projects, DFG, the Next Generation
Infrastructure (Bsik), Bi-national US-Israel Science Foundation (BSF), and the Israel Science Foundation for financial
support.

\bibliographystyle{rQUF}


\label{lastpage}

\end{document}